\documentstyle[amssymb,prl,aps,twocolumn,epsf]{revtex}

\begin{document}
\title{Double Quantum Dots as Detectors of High-Frequency Quantum Noise in
Mesoscopic Conductors. }
\author{Ram\'{o}n Aguado$^{1,2}$ and Leo P. Kouwenhoven$^{1}$}
\address{1-Department of Applied Physics and DIMES, Delft University of Technology,
P.O.Box 5046, 2600 GA, Delft, The Netherlands.\\
2-Department of Physics and Astronomy, Rutgers University, Piscataway, NJ
08854-8019, USA.}
\maketitle

\begin{abstract}
We propose a measurement set-up for detecting quantum noise over a wide
frequency range using inelastic transitions in a tunable two-level system as
a detector. The frequency-resolving detector consists of a double quantum
dot which is capacitively coupled to the leads of a nearby mesoscopic
conductor. The inelastic current through the double quantum dot is
calculated in response to equilibrium and non-equilibrium current
fluctuations in the nearby conductor, including zero-point fluctuations at
very low temperatures. As a specific example, the fluctuations across a
quantum point contact are discussed.
\end{abstract}

\pacs{PACS numbers:XXX}

\tighten

\thispagestyle{empty}

\narrowtext
Two-level systems (TLS) coupled to a dissipative environment are canonical
model systems to study dephasing in quantum mechanics\cite{Cal2}. The
reversed problem is a TLS that measures the characteristics of a specific
environment. The transition rate for levels separated by an energy $\epsilon 
$, is a measure of the spectral density of the fluctuations in the
environment at a frequency $f=\epsilon /h$. Transitions are allowed when
energy can be exchanged with the environment. Recently, two device
structures were realized that can be used as tunable TLS. In a
superconducting single-electron transistor a Cooper-pair \cite{JQP} and in a
double quantum dot (DQD) an electron \cite{Tosh} can make inelastic
transitions between two discrete energy states. In this work we calculate
the rate for inelastic transitions in a DQD coupled to an environment formed
by a second mesoscopic device.

Small electronic devices have interesting equilibrium and non-equilibrium
noise properties which are non-linear in frequency \cite{Kog}. In
equilibrium, a transition occurs going from low-frequencies, where
Johnson-Nyquist noise due to thermal fluctuations dominates, to high
frequencies where quantum noise due to zero-point fluctuations (ZPF)
prevails. When the device is voltage biased, non-equilibrium fluctuations
can become dominant. These lead to shot noise in the current, which has been
measured near zero-frequency \cite{Been} and at several high frequency
values where ZPF become dominant.\cite{Scho} The idea of using a
mesoscopic device --quantum point contact (QPC)-- as an environment for
another device --quantum dot-- has successfully been used in the so-called
'which-path' detector: \cite{Bucks} the dc shot-noise of the QPC modifies
the transport properties of the dot, leading to dephasing.\\
Here we propose a setup for studying the effect of {\it broad-band
fluctuations} on the inelastic rate in a TLS. This setup provides a {\it %
frequency-resolved} detection over a large frequency range of the
fluctuations in mesoscopic 
systems. A wide
frequency range requires that the frequency dependent impedance of the whole
circuit is taken into account. Below, we first 
describe the basic properties of a DQD, then formulate transition
probabilities in terms of the noise spectrum, followed by calculations where
the specific environment is formed by a QPC.

\begin{figure}[h]
\centerline{\epsfxsize=0.45\textwidth
\epsfbox{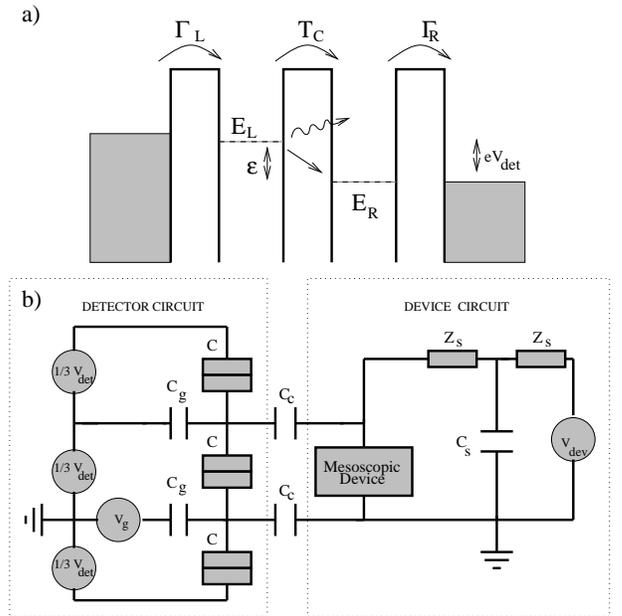}
}
\caption{(a) Energy diagram of a DQD in the regime of high bias voltage. (b)
Circuit for capacitively coupling the DQD to a second mesoscopic device,
e.g. a QPC. The detector and device circuits are separately biased by
different voltages. The symbols $\boxminus $ in the detector correspond to
the three tunnel barriers in a DQD.}
\label{fig:schematic1}
\end{figure}
A DQD is a fully-controllable TLS. The separation
between levels $\epsilon \equiv E_{L}-E_{R}$, the tunnel rates across the
left and right barriers, $\Gamma _{L}$, $\Gamma _{R}$, and the tunnel
coupling between the dots, $T_{c}$ (see Fig.1a), can be tuned separately by
means of gate voltages. If $\epsilon \gg T_{c}$, we can neglect coherence
effects due to mixing between $E_{L}$ and $E_{R}$ \cite{Oos}. Then a
non-zero current for $\epsilon \neq 0$ necessarily involves emission ($%
\epsilon >0$) or absorption ($\epsilon <0$) of quanta to or from the
environment. An applied bias voltage, $V_{\det }$, shifts the two Fermi
levels in the two leads. The higher Fermi energy in the left lead allows
that for $\epsilon >0,$ the high energy state, $E_{L},$ can be occupied by
tunneling through the left barrier. 
An inelastic transition with rate $\Gamma _{i}$, followed by tunneling
through the right barrier yields an inelastic current, $I_{inel}(\epsilon )=%
\frac{e}{\hbar }(\Gamma _{L}^{-1}+\Gamma _{i}^{-1}+\Gamma _{R}^{-1})^{-1}$.
When the central tunnel barrier is made the largest, i.e. $\Gamma
_{i}<<\Gamma _{L},\Gamma _{R}$, the inelastic current is governed by time
dependent fluctuations $\delta \epsilon (t)=e\delta
V(t)$ and can be calculated from perturbation theory on $T_{c}$:\cite
{Cal2,Gla} 
\begin{eqnarray}
I_{inel}(\epsilon )\simeq \frac{e}{\hbar }\Gamma _{i}(\epsilon )=\frac{e}{%
\hbar }T_{c}^{2}P(\epsilon )
\end{eqnarray}
$P(\epsilon )$ being the the probability for the exchange of energy quanta
with the environment:\cite{Dev,Gir,Yul} 
\begin{eqnarray}
P(\epsilon )=\frac{1}{2\pi \hbar }\int_{-\infty }^{\infty }exp[J(t)+i\frac{%
\epsilon }{\hbar }t]dt
\end{eqnarray}
All the information about the environment is contained in the
autocorrelation of the phase operators 
$J(t)\equiv \langle [\delta \hat{\phi} (t)-\delta \hat{\phi} (0)]\delta
\hat{\phi} (0)\rangle $. \cite{gauss} $\delta \hat{\phi} (t)=\frac{e}{\hbar }\int^{t}dt^{\prime
}\delta \hat{V}(t^{\prime })$ are the conjugate phases of the voltage
fluctuations $\delta \hat{V}(t)$ (characterized by the 
spectral density $S_{V}(\omega )$). We are interested in current
fluctuations, $S_{I}(\omega )$, generated in a nearby mesoscopic device.
Fig. 1(b) shows an example of a circuit that capacitively couples a
mesoscopic device to a DQD. Both noise spectra are related through: $%
S_{V}(\omega )=|Z(\omega )|^{2}S_{I}(\omega )$ ($Z(\omega )$ being the
trans-impedance connecting detector and device circuits) and then:
\begin{eqnarray}
J(t)=\frac{2\pi }{\hbar R_{K}}\int_{-\infty }^{\infty }\frac{|Z(\omega )|^{2}%
}{\omega ^{2}}S_{I}(\omega )(e^{-i\omega t}-1)d\omega,
\end{eqnarray}
$R_{K}=h/e^{2}\simeq 25.8k\Omega $ is the quantum resistance.
Importantly, $S_{I}(\omega )\equiv \int_{-\infty }^{\infty }d\tau e^{i\omega
\tau }\langle \delta \hat{I}(\tau )\delta \hat{I}(0)\rangle $ appears in a {\it %
non-symmetrized} form.\cite{nonsym}
We shall demonstrate that this is crucial to account
correctly for ZPF. Eqs. 1-3 relate the inelastic current through
the DQD to the noise spectrum of an arbitrary, nearby mesoscopic device.\cite
{phon}

Our problem is now reduced to the determination of $Z(\omega )$ and $%
S_{I}(\omega )$ for a specific device embedded in a specific circuit. We
consider the circuit in Fig. 1b for coupling the current fluctuations from
the device via the capacitors $C_{c}$ into the DQD. The DQD is modelled as
three tunnel barriers with capacitances $C$ and biased by a voltage $V_{det}$%
. The gate voltage, $V_{g},$ controls $\epsilon $. The device is connected
to a voltage source, $V_{dev},$ via leads characterized by the impedances $%
Z_{s}$ and the capacitor $C_{s}$. For this circuit we obtain for the
trans-impedance: 
\begin{eqnarray}
Z(\omega )=\frac{\alpha _{1}Z_{s}}{(\alpha _{2}[1+i\omega Z_{s}C_{c}-\frac{1%
}{(2+i\omega Z_{s}C_{S})}]-\alpha _{3}i\omega Z_{s}C_{c})}
\end{eqnarray}
with $\alpha _{1}=\frac{C+C_{g}+C_{c}}{C}$, $\alpha _{2}=\frac{%
(2C+C_{g}+C_{c})^{2}-C^{2}}{C_{c}C}$ and $\alpha _{3}=\frac{2C+C_{g}+C_{c}}{C%
}$. For small $\omega$, $|Z(\omega )|^{2}\simeq \rho [\gamma /(\gamma
^{2}+\omega ^{2}]$ with $\rho =\alpha _{1}^{2}Z_{s}/[\alpha _{2}(\alpha
_{2}(C_{c}+C_{s})-\alpha _{3}C_{c})]$ and $\gamma =\alpha _{2}/[Z_{s}(\alpha
_{2}(C_{c}+C_{s})-\alpha _{3}C_{c})]$. At $\omega=0$, it reaches the maximum
value $|Z(0)|^{2}=\alpha _{1}^{2}/\alpha _{2}^{2}Z_{s}^{2}\equiv\kappa
^{2}R_{K}^{2}$ (i.e. Ohmic environment). 
\begin{figure}[h]
\centerline{\epsfxsize=0.45\textwidth
\epsfbox{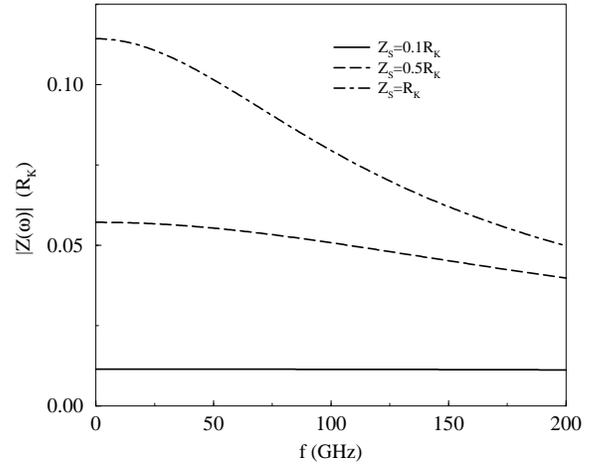}
}
\caption{Frequency dependence of $|Z(\omega )|$ for different Ohmic
resistors in the leads. We have taken typical experimental values for: $%
C=0.05fF,$ $C_{g}=C_{c}=0.1fF,$ $C_{s}=1nF$.}
\label{fig:schematic1}
\end{figure}
In Fig. 2 we plot $|Z(\omega )|$ for different values of $Z_{s}$ and typical
experimental values for the elements in the circuit. For $%
Z_{s}\rightarrow 0$ the device is shorted (i.e. $Z(\omega )\rightarrow 0$)
and the detector is insensitive to the fluctuations. For{\bf \ }$%
Z_{s}<0.1R_{K}$, the trans-impedance is approximately independent of
frequency and can be written as $|Z(\omega )|^{2}\simeq |Z(0)|^{2}$. In this
case, $I_{inel}$ is determined only by the frequency dependence of
the noise and therefore easier to interpret. If $Z_{s}=0.1R_{K}$ is taken,
then the coupling of the noise into the detector is sufficiently effective.
Provided that $J(t)$ does not diverge for long times (see below) we expand $%
e^{J(t)}\simeq 1+J(t)$ in Eq. 2 and derive:
\begin{eqnarray}
P(\epsilon ) &\simeq& \{1-\frac{2\pi }{\hbar R_{K}}\int_{-\infty }^{\infty
}d\omega \frac{|Z(\omega )|^{2}}{\omega ^{2}}S_{I}(\omega )\}\delta
(\epsilon )\nonumber\\ 
&+&\frac{2\pi }{R_{K}}\frac{|Z(\epsilon /\hbar )|^{2}}{\epsilon ^{2}}%
S_{I}(\epsilon /\hbar ).
\end{eqnarray}
The first part renormalizes the elastic current (i.e. when $\epsilon =0$),
which we do not consider further here. Inserting the last term in Eq. 1 we
obtain the inelastic current through the DQD detector: 
\begin{eqnarray}
I_{inel}(\epsilon )\simeq 4\pi ^{2}\kappa ^{2}\frac{T_{c}^{2}}{e}\frac{%
S_{I}(\epsilon /\hbar )}{\epsilon ^{2}}.
\end{eqnarray}
We note that current fluctuations at frequency $\omega $ result in an
inelastic current at level difference $\epsilon =\hbar \omega $. Below we
discuss that this detector current is asymmetric in the absorption ($%
\epsilon <0$) and emission ($\epsilon >0$) sides, which results from the
asymmetry in the noise due to ZPF.

As an application, we study the current noise spectrum of a QPC. 
The right quantity to be calculated for our purposes is the {\it %
non-symmetrized} noise. The symmetrized version
has been calculated in Refs.\onlinecite {Les,Yan,But}. 
The time dependent fluctuations of the current around its average
are
$\delta \hat{I}(\tau )=\hat{I}(\tau )-\langle \hat{I}(\tau )\rangle $, 
the  current operator being
$\hat{I}(t)=\frac{2e}{h}\sum_{\alpha ,\beta }\int\int_{-\infty }^{\infty
}d\epsilon _{1}d\epsilon _{2}I_{\alpha ,\beta
}(\epsilon _{1},\epsilon _{2})\hat{a}_{\alpha ,\epsilon _{1}}^{\dagger }(t)%
\hat{a}_{\beta ,\epsilon _{2}}(t)$
($\hat{a}_{\alpha ,\epsilon }^{\dagger }(t)$ $[\hat{a}_{\alpha ,\epsilon
}(t)]$ is the creation [annihilation] Heisenberg operator
of the scattering state $\psi _{\alpha }(\vec{r},\epsilon )$,
$\alpha \equiv (a,n)$ and $\beta \equiv (b,m)$
represent summations over leads and number of channels and
$I_{\alpha ,\beta }(\epsilon _{1},\epsilon _{2})$
are the matrix elements of the current with
respect to these scattering states).
The non-symmetrized noise spectrum can be written as:
$S_{I}(\omega )\equiv \int_{-\infty }^{\infty }d\tau e^{i\omega
\tau }\langle \delta \hat{I}(\tau )\delta \hat{I}(0)\rangle 
=\frac{4e^{2}}{h}\sum_{a,b,n,m}\int d\epsilon
I_{a,n,b,m}I_{b,m,a,n}f_{a}(\epsilon )[1-f_{b}(\epsilon +\hbar \omega )]$;
$f(\epsilon )$ being the Fermi-Dirac function.
If the matrix elements are expressed in
terms of energy-independent transmission and reflection scattering matrices,
we obtain (see \cite{Been} for a
complete derivation in the $\omega =0$ limit): 
\begin{eqnarray}
S_{I}(\omega ) &=&\frac{4}{R_{K}}\sum_{m}^{N}D_{m}(1-D_{m})\{\frac{%
(eV_{dev}+\hbar \omega )}{1-e^{-\beta (eV_{dev}+\hbar \omega )}}  \nonumber
\\
&+&\frac{(\hbar \omega -eV_{dev})}{1-e^{-\beta (\hbar \omega -eV_{dev})}}\}+%
\frac{4}{R_{K}}\sum_{m}^{N}D_{m}^{2}\frac{2\hbar \omega }{1-e^{-\beta \hbar
\omega }},
\end{eqnarray}
where $N$ is the number of channels, $D_{m}$ is the
transmission probability of the $m^{th}$ channel, $V_{dev}
$ is the applied voltage and $\beta =1/k_{B}T$. In equilibrium (i.e. $%
V_{dev}=0$), we recover the fluctuation-dissipation theorem \cite{Call} $%
S_{I}(\omega )=2G\frac{2\hbar \omega }{1-e^{-\beta \hbar \omega }},$ where $%
G=\frac{2}{R_{K}}\sum_{m}^{N}D_{m}$ is the conductance. Here, our model
reduces to the usual theory for the effects of an electromagnetic
environment on single electron tunneling \cite{Dev,Gir,Yul}. Equation (7) is
not symmetric ($S_{I}(\omega )>S_{I}(-\omega )$) which results from
the difference between emission and absorption due to ZPF.
This has to be compared with the symmetrized version 
\onlinecite{Les,Yan,But} where $S_{I}(\omega )=S_{I}(-\omega )$.\newline
From now on we present calculations for zero temperature.
Fig. 3a shows the noise, 
$S_{I}(\nu ),$ vs. the normalized frequency 
$\nu =\frac{\hbar \omega }{|eV_{dev}|}$, for $N=2$ and different values for $%
D=\sum_{m}^{N}D_{m}$. For $\nu >0$ the noise increases linearly with
frequency (with a slope determined by $D$) due to ZPF. For $\nu <0$ there
are two different cases: for non-open channels the non-equilibrium
shot-noise dominates when $-1<\nu <0,$ whereas for $\nu <-1$ the noise is
zero. For open channels ($D$ = 1 and $D$ = 2) the noise
is always zero on the absorption side.

The inset to Fig. 3a shows the voltage dependence of the noise for different
values of $\omega $ and $D$ = 1.5 (see also Ref. \cite{Scho}). When $%
|eV_{dev}|<\hbar \omega $ the noise spectrum is flat. Here, the fluctuations
are dominated by quantum noise and do not change from the equilibrium value
for small voltages. For $|eV_{dev}|>\hbar \omega $ the noise increases
linearly with the voltage and is due to shot noise. This transition from
quantum to shot noise can directly be tested by measuring the detector
current at a fixed level separation, $\epsilon =\hbar \omega ,$ as a
function of the voltage across the QPC.

In Fig. 3b we plot $I_{inel}(\nu )$ for the same values
of the total transmission. The main feature is an asymmetric broadening in
the emission ($\nu >0$) and absorption ($\nu <0$) sides. We first consider
the absorption side. For open channels ($D$ = 1 and $D$ = 2),
the nonequilibrium part of the noise is zero and no energy can be absorbed
by the detector. For non-open channels ($D$ = 1.5), the
nonequilibrium noise is finite and the detector can absorb energy even at
zero temperature; this is reflected as an inelastic current through the DQD
for $\nu <0$. Emission is possible for both open and non-open channels due
to ZPF so the inelastic current for $\nu >0$ is always
finite. 
\begin{figure}[h]
\centerline{\epsfxsize=0.45\textwidth
\epsfbox{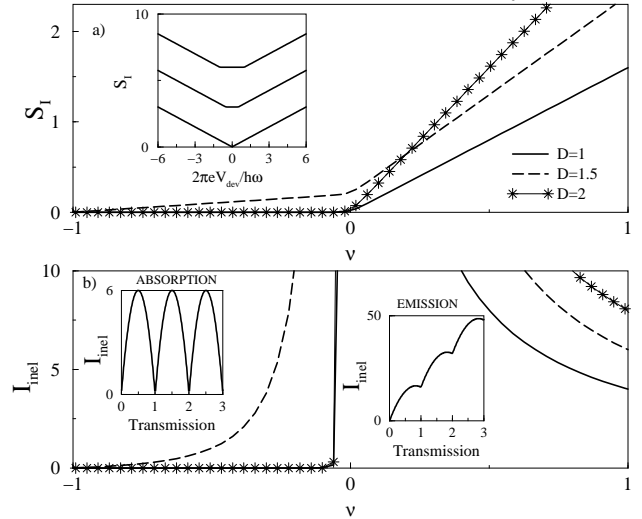}
}
\caption{ (a) $S_{I}(\nu )$ (units of $\frac{4e^{2}}{h}eV_{dev}$) for
different values of $D$. Inset: $S_{I}$ (units of $\frac{2e^{2}}{\pi}\omega$%
) vs. $\frac{eV_{dev}}{\hbar }$ (units of $\omega $) for different $\omega
=1,0.5,0$, from top to bottom. (b) $I_{inel}(\nu )$ (units of $32\pi
^{2}\kappa ^{2}\frac{e}{h}\frac{T_{c}^{2}}{eV_{dev}}$). Insets: transmission
dependence of $I_{inel}$ for a fixed value of the frequency; $\nu =-0.25$
for absorption and $\nu =0.25$ for emission. }
\label{fig:noise}
\end{figure}
The transmission dependence is shown in the insets to Fig. 3(b). At fixed $%
\nu$ the absorption oscillates as a function of $D$ (left inset) whereas the
emission is an increasing function with plateau-like features (right inset).

To check the observability of these predictions we take as an example $D=1.5$%
, $\nu =0.5$, $\kappa =10^{-2}$ and $T_{c}=10\mu eV$. This implies, e.g. for
the absorption side $I_{inel}$ $\simeq 12pA$ at $\epsilon /h=1.22GHz$ ($%
eV_{dev}=10\mu eV$) For $\epsilon /h=12.17GHz$ ($eV_{dev}=100\mu eV$), $%
I_{inel}\simeq 1.2pA$. These values are well within the resolution limits of
present day techniques.\cite{Tosh,Scal}\newline
A log-log plot of $I_{inel}$ vs. $\nu $ (Fig 4) demonstrates the transition
from quantum to shot noise. At $\nu =-1$ (absorption), a sharp decline of
the current marks such a transition. For open channels, the current on the
emission side follows a power law behaviour indicating the occurence of an
infrared divergence (see below).\newline
We now examine the validity of our previous results. For 
non-open channels, $J(t)$ behaves for long times as:
\begin{eqnarray}
J(t\rightarrow \infty )\simeq -8\pi ^{2}\kappa
^{2}\sum_{m}^{N}D_{m}(1-D_{m})|\frac{eV_{dev}}{\hbar }t|.
\end{eqnarray}
This linear time dependence coincides with previous results at equilibrium
for an ohmic environment at finite temperature \cite{Dev,Gir,Yul}. Eq. 8
defines two regimes: (a) $\nu >\nu _{c}$ with $\nu _{c}=8\pi ^{2}\kappa
^{2}\sum_{m}^{N}D_{m}(1-D_{m})$, the expansion for $J(t)$ in Eqs. 5-6 is
valid. (b) $\nu <\nu _{c},$ the expansion for $J(t)$ cannot be used. Here,
we obtain from Eq. 8: 
$I_{inel}(\epsilon )\simeq \frac{eT_{c}^{2}}{\pi \hbar }\frac{\eta }{\eta
^{2}+\epsilon ^{2}}$
with $\eta =eV_{dev}\nu _{c}/8$ \cite{heat}. Our results are within the
limits of validity of regime (a). 
\begin{figure}[tbp]
\centerline{\epsfxsize=0.40\textwidth
\epsfbox{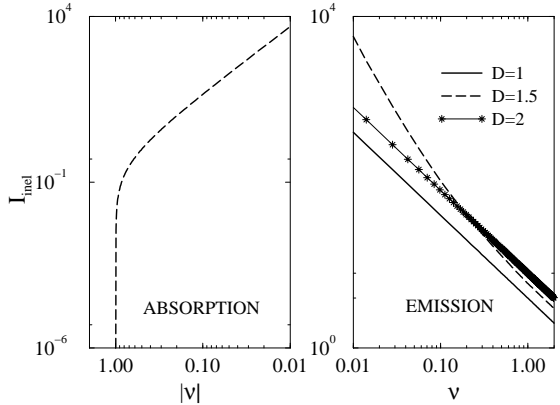}
}
\caption{$I_{inel}(\nu )$ vs. $\nu $ (log-log). Same units as Fig. 3b.}
\label{fig:noise}
\end{figure}
Note that for open channels 
$J(t\rightarrow \infty )\simeq -\lambda \{ln|\frac{eV_{dev}}{\hbar }t|+\xi +i%
\frac{\pi }{2}\}$
where $\xi =0.5772...$ is Euler's constant and $\lambda =8\pi \kappa
^{2}R_{K}G$. This behaviour leads to the infrared divergence $P(\epsilon
)\simeq \epsilon ^{\lambda -1}$ caused by the ZPF of the electron-hole pair
excitations in the QPC.\cite{Xray}\newline
In conclusion, the inelastic current through a DQD at low temperatures can
provide a broad-band frequency resolved measurement of the equilibrium and
non-equilibrium fluctuations in a nearby mesoscopic conductor. The asymmetry
between absorption and emission processes gives a clear measurement of the
non-equilibrium quantum noise. The predicted signal is well within the
resolution limits of present day experiments on quantum dots \cite{Tosh} as
well as on superconducting circuits \cite{JQP}. In the present case, the
measurement by the DQD has a negligible effect on the transport through the
QPC. If the QPC is replaced by a circuit in which superposition of quantum
states is important (e.g. strongly coupled quantum dots), then our
detection setup forms an interesting quantum measurement 
problem. \cite{Which}\newline
We acknowledge Yuli Nazarov for fruitful criticism and
discussions. We also thank Cees Harmans, Leonid Glazman, Wilfred van der
Wiel, Toshimasa Fujisawa, Tjerk Oosterkamp, Michael Janus, Yann Kervennic,
Silvano De Franceschi and Leonid Gurevich. Work supported by:
the Dutch Organization for Fundamental Research on Mater (FOM), the NEDO
joint research program (NTDP-98), the EU via a TMR network, the NSF
grant DMR 97-08499 and the MEC of Spain grant PF 98-07497938.


\begin{references}
\bibitem{gauss}  This is valid for gaussian fluctuations \cite{Yul} 
provided the following
realistic
experimental conditions are met: a) weak coupling between detector and fluctuations
and measurements in the linear response regime of the device.
As long as tunneling events in the DQD weakly excite the environment only the power spectrum
enters in $J(t)$. b) large level separation: for low frequencies a weakly perturbed environment
is no longer well described by a set of harmonic oscillators.
\bibitem{phon}  Our formulation does not include other degrees of freedom of
the environment (phonons, for example). At very low temperatures, and in the
absence of any extra device, $I_{inel}(\epsilon >0)$ is finite due to
spontaneous emission.\cite{Tosh} $I_{inel}(\epsilon <0)$ (absorption side),
however, is virtually zero. This makes the absorption region ideal for
measuring non-equilibrium fluctuations. The different sources of the noise
in the emission side can be separated by studying the dependence on
different parameters of the device.
\bibitem{Scal}  $I_{inel}$ scales with $eV_{dev}$ at fixed $\nu $, which can
be used to distinguish from heating effects.
\end{references}
\end{document}